\documentclass[10pt,prd,longbibliography,superscriptaddress]{revtex4-2}
\usepackage[T1]{fontenc}
\usepackage[utf8]{inputenc}
\usepackage[english]{babel}
\usepackage{gensymb}

\usepackage{amsfonts,amsbsy,amssymb,amsmath,graphicx,float} 
\usepackage{color}

\graphicspath{{figures/}}

\DeclareUnicodeCharacter{2009}{\_}

\begin{document}

\title{The Nature of Reactive and Non-reactive Trajectories for a Three Dimensional Caldera Potential Energy Surface}

\author{Matthaios Katsanikas}
\email{mkatsan@academyofathens.gr}
\affiliation{Research Center for Astronomy and Applied Mathematics, Academy of Athens, Soranou Efesiou 4, Athens, GR-11527, Greece.}
\affiliation{School of Mathematics, University of Bristol, \\ Fry Building, Woodland Road, Bristol, BS8 1UG, United Kingdom.}
\author{Stephen Wiggins}
\affiliation{School of Mathematics, University of Bristol, \\ Fry Building, Woodland Road, Bristol, BS8 1UG, United Kingdom.}

\begin{abstract}
We used for the first time the method of periodic orbit dividing surfaces in a non-integrable Hamiltonian system with three degrees of freedom. We have studied the structure of these four dimensional objects in the five dimensional phase space. This method enabled us  to detect the reactive and non-reactive trajectories  in a three dimensional Caldera potential energy surface. 
We distinguished four distinct types of trajectory behavior. Two of the types of trajectories could only occur in a three dimensional Caldera potential energy surface and not in a two dimensional surface, and we have shown that this is a result of homoclinic intersections. These homoclinic intersections were analyzed with the method of Lagrangian descriptors. Finally, we were able to detect and describe the phenomenon of dynamical matching in the three dimensional Caldera potential energy surface, which is an important mechanism for understanding the reaction dynamics of organic molecules.
\end{abstract}

\maketitle

\noindent\textbf{Keywords:} Periodic orbit dividing surfaces, Caldera Potential, Dynamical matching, Chemical reaction dynamics.

\section{Introduction}
\label{intro}

Dynamical matching, originally observed by Carpenter \cite{carpenter1995,carpenter1985} ,  is an interesting phenomenon that is playing an increasingly important role for understanding the reactions of organic molecules, such as, the  vinylcyclopropane-cyclopentene rearrangement \cite{baldwin2003,gold1988}, the stereomutation of cyclopropane \cite{doubleday1997}, the degenerate rearrangement of bicyclo[3.1.0]hex-2-ene \cite{doubleday1999,doubleday2006} or that of 5-methylenebicyclo[2.1.0]pentane \cite{reyes2002}. 

Carpenter observed that the dynamical matching mechanism could be described by a two dimensional potential energy surface (PES) that resembles the collapsed region of an erupted volcano, to which the name ‘’caldera’’ was given in \cite{doering2002}. In particular, the caldera PES originally described by Carpenter  has one minimum at the center and is surrounded by  four index-1 saddles  that control the entrance into and exit from the caldera region. Two of these index-1 saddles  have low values of energy and the other two have high energy values, with respect to each other.  The  lower index-1 saddles correspond to the formation of the chemical products and the two higher index-1 saddles correspond to the reactants.

Trajectory behavior in this two dimensional (2D) Caldera PES has been extensively studied in recent years, see \cite{collins2014,katsanikas2018,katsanikas2019,katsanikas2020a,katsanikas2020b,Agaoglou2019}. A crucial requirement of these studies is to obtain collections of initial conditions corresponding to trajectories whose past and future behavior can be monitored in terms  of the topographical features of the Caldera PES.  For the associated two degree-of-freedom Hamiltonian systems this is accomplished with the method of periodic orbit dividing surfaces (\cite{Pechukas73,Pechukas77,Pechukas79,pechukas1981,Pollak78,pollak1985}. The dividing surfaces obtained from this method have one less dimension than the energy surface  and are constructed from a periodic orbit.  They satisfy the no-recrossing property (see for example \cite{ezra2018}).  If we consider trajectories that are initiated on the dividing surfaces of the  unstable periodic orbits associated with the  higher energy  index-1 saddles we have two cases for the trajectory behavior(see \cite{katsanikas2018,katsanikas2019,katsanikas2020a,katsanikas2020b}):

\begin{enumerate}

    \item In the first case, the trajectories that are initiated on the dividing surfaces of the unstable periodic orbits of the higher energy  index-1 saddles go straight across the caldera and exit through the region of the opposite lower energy saddle. This trajectory behavior is what is referred to as {\em dynamical matching}. Understanding this behavior is very important for the caldera-type chemical  reactions since 100$\%$ of the trajectories entering the Caldera from a region of one of the higher energy index-1 saddles exit through the region of  one index-1 saddle, instead of $50\%$ crossing each of the lower energy index-1 saddles  which is the prediction from statistical theories (see more details in the introduction of \cite{collins2014} and \cite{carpenter1985,carpenter1995}). We have previously shown that this behavior always occurs in  2D symmetric caldera potential energy surfaces because of the non-existence  of intersections  between the unstable invariant manifolds of the unstable periodic orbits of the higher index-1 saddles with the stable invariant manifolds of the unstable periodic orbits of the central area of the Caldera (see \cite{katsanikas2018,katsanikas2020a}). If we break the symmetry, we can observe the trajectory behavior of the second case.

 \item In the second case, the trajectories that start from the region of the higher energy  index-1 saddles enter  the caldera and may become  trapped for a period of time before they exit from one of the regions of  any of the four index-1 saddles. This is because the trajectories that have initial conditions on the dividing surfaces of the unstable periodic orbits of the higher energy index-1 saddles are trapped in the lobes between the unstable invariant manifolds of the unstable periodic orbits of the higher energy index-1 saddles and the stable invariant manifolds of unstable periodic orbits that exist in the central area of the Caldera (see \cite{katsanikas2020a,katsanikas2019}). This case happens when we stretched the potential in the  horizontal direction (see \cite{katsanikas2019,katsanikas2020a,katsanikas2020b} or after a bifurcation of critical points (\cite{geng2021influence,geng2021bifurcations} (in which we have a transition from a Caldera potential energy surface from one central minimum and four index-1 saddles to a Caldera with three index-1 saddles).
    
\end{enumerate}

In this paper we develop a 3D model of the Caldera PES. This 3D extension of the 2D caldera potential energy surface is obtained by coupling a  harmonic oscillator to one of the configuration space coordinates of the 2D Caldera PES. Hence, one can view this as a simple model of the 2D Caldera PES in a bath (with only a single bath mode). 

We construct a 3D model of the Caldera PES and analyze the trajectory behavior. However, the passage to 3 dimensions brings significant new difficulties when attempting a similar analysis as the 2D case.  In particular, the  classical 2 DoF construction of  periodic orbit dividing surfaces  (\cite{Pechukas73},\cite{Pollak78}) fails. The reason is using this approach the periodic orbits (1 dimensional objects) does not result in a 4 dimensional dividing surfaces in the 5 dimensional energy surface, which is what is required for a 3 DoF system. 

The construction of dividing surfaces in Hamiltonian systems with three, and more,  degrees of freedom has been accomplished with the use of a normally hyperbolic invariant manifold -NHIM (\cite{wiggins1994},\cite{wiggins2001},\cite{komatsuzaki2003},\cite{uzer2002}\cite{wiggins2016}). But the computation of this object is difficult and limited in many cases and it requires extensive numerical calculations. Recently, we have  developed a method that generalizes the periodic orbit dividing surface construction in Hamiltonian systems with three or more degrees of freedom (\cite{katsanikas2021ds},\cite{katsanikas2021dsa}) that does not require the computation  of a NHIM.  In this paper we show how our construction of periodic orbit dividing surfaces can be applied in 3D for the analysis of dynamical matching.

This paper is outlined as follows. In Section \ref{model} we describe our model of a 3D Caldera PES. The results are described in Section \ref{results}. In particular, in Section \ref{sub1} we describe the construction of the 3D periodic orbit dividing surface and in Section \ref{sub2} we describe our analysis of trajectories that enter the Caldera. In Section \ref{conc} we describe our conclusions.

\section{Model}
\label{model}

The 3D potential energy surface (PES)  for our model is constructed  from  a familiar 2D Caldera potential energy surface (that is symmetric with respect to the $y$-axis) and a  harmonic oscillator in the $z$-direction that is coupled with the symmetry axis ($y$-axis) of the 2D Caldera PES. This can be viewed as a 2D Caldera coupled to a bath, where the bath has a single mode.

The 2D caldera PES that we used for our potential (see \cite{collins2014})  has the form:

\begin{equation}
    \begin{aligned}
    V_c(x,y) &= c_1r^2+c_2y-c_3r^4\cos(4\theta)\\ &= c_1(x^2+y^2)+c_2y-c_3(x^4+y^4-6x^2y^2),
    \end{aligned}
\end{equation}

\noindent
where $(x,y)$  are Cartesian coordinates, $(r,\theta)$  are the standard polar  coordinates, and  $c_1,c_2,c_3$ are parameters. We used the same values for the parameters as in \cite{collins2014},\cite{katsanikas2018},\cite{katsanikas2019},
\cite{katsanikas2020a} and \cite{katsanikas2020b}, i.e.  $c_1=5,c_2=3,c_3=-0.3$.

The 3D PES of our system, the 2D caldera potential plus a coupled harmonic oscillator in the $z$-direction, has the form:

\begin{equation}
    \begin{aligned}
    V(x,y,z) &=V_c(x,y)+c_4 z^2 + c_5 yz^2,
    \end{aligned}
\end{equation}

\begin{figure}
 \centering
 \includegraphics[scale=0.73]{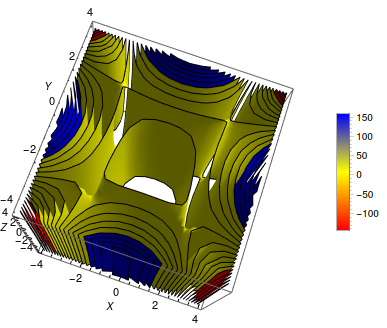}\\
\caption{The contours of the 3D caldera potential in the $(x,y,z)$ space (where the color represents  the value of the potential energy).}
\label{3d-potential}
\end{figure}

\begin{figure}
 \centering
A)\includegraphics[scale=0.55]{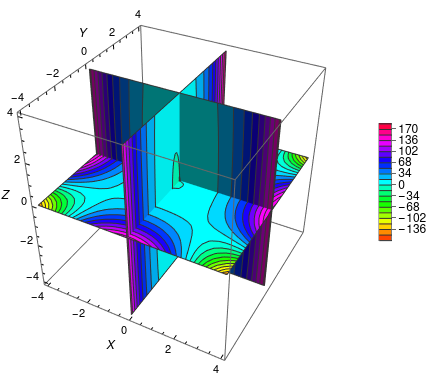}
B) \includegraphics[scale=0.55]{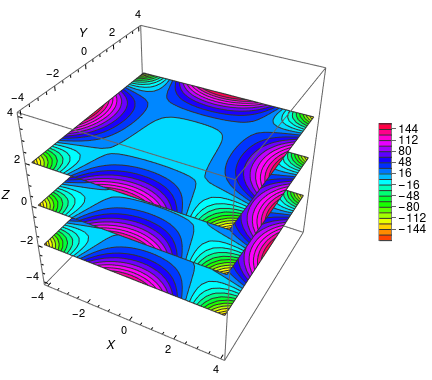}\\
C)\includegraphics[scale=0.55]{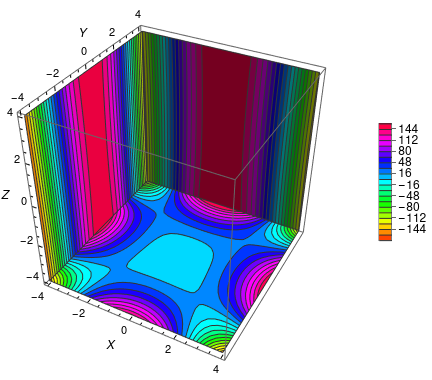}
D) \includegraphics[scale=0.55]{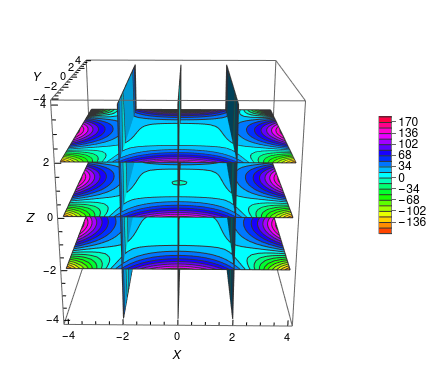}\\
\caption{The slices of the  contours of the 3D caldera potential in the $(x,y,z)$ space. A) The slices for $x=0$ and $y=0$. B) The slices for $z=-2$, $z=0$ and $z=2$. C) The slices for $x=-4$ and $y=4$. D) The slices for  $z=-2$, $z=0$, $z=2$, $x=-2$, $x=0$ and $x=-2$. }
\label{3d-potential1}
\end{figure}

\noindent
Where $c_4z^2$ and $c_5yz^2$ represents  the harmonic oscillator and the coupling term, with $c_4=0.2$ and $c_5=0.03$. For a geometrical representation of this potential we show the 3D contours of this potentials in Fig. \ref{3d-potential} or the contours  of particular slices of this potential in Fig. \ref{3d-potential1}. We see that this potential has the same topography in the $(x,y)$ space as that of the  2D Caldera, as can be seen from  panels C  and D of Fig. \ref{3d-potential1}. We observe that the potential walls that are around the caldera  in the $(x,y)$ subspace of the configuration space continue into the $z$-direction (Fig. \ref{3d-potential} and panel A of Fig. \ref{3d-potential1}). Furthermore, the topography of the potential for different slices in the $z$-direction is similar to that of the potential on the plane $z=0$ (that is equivalent to the potential of the 2D caldera). 

The 3 degree-of-freedom (DoF) Hamiltonian is given by:

\begin{equation}
    H(x,y,z,p_x,p_y,p_z)=\frac{p_x^2}{2m}+\frac{p_y^2}{2m}+\frac{p_z^2}{2m}+V(x,y,z),
\end{equation}

\noindent
 where $(x,y,z)$  are Cartesian coordinates and $p_x$, $p_y$, $p_z$  denote the momentum of the particle in $x-$, $y-$ and $z-$ direction respectively. We consider $m$ to be a constant equal to $1$. The Hamiltonian equations of motion are therefore:

\begin{equation}
    \begin{aligned}
    &\dot{x} = \frac{\partial H}{\partial p_x} = \frac{p_x}{m}\\
    &\dot{y} = \frac{\partial H}{\partial p_y} = \frac{p_y}{m}\\
    &\dot{y} = \frac{\partial H}{\partial p_z} = \frac{p_z}{m}\\
    &\dot{p_x} = - \frac{\partial V}{\partial x}(x,y,z) = -(2c_1x-4c_3x^3+12c_3xy^2)\\
    &\dot{p_y} = - \frac{\partial V}{\partial y}(x,y,z) = -(2c_1y-4c_3y^3+12c_3x^2y+c_2+c_5z^2)\\
    &\dot{p_z} = - \frac{\partial V}{\partial y}(x,y,z)=-(2c_4z+2c_5yz)\\
    \end{aligned}
\end{equation}

The 2D Caldera as we described in the introduction has one central minimum surrounded by 4 index-1 saddles. The  3D Caldera has the same equilibrium points (with $z=0$ for every equilibrium point). We give the position and the value of energy of the equilibrium points of the 3D Caldera potential in table I (for the set of the parameters $c_1,c_2,c_3,c_4$ and $c_5$ that we used in our paper).

\begin{table}
\begin{center}
\caption{Stationary points of the Caldera potential for $c_1=5,c_2=3$  $c_3=-0.3,c_4=0.2$ and $c_5=0.03$  ("RH" and "LH" are the abbreviations for right hand and left hand respectively)}
\begin{tabular}{l  l  l  l l}
\hline
Critical point & x & y & z & E \\
\hline
Central minimum & 0.000 & -0.297 & 0 & -0.448 \\
Upper LH saddle  & -2.149 & 2.0778 & 0 & 27.0123 \\
Upper RH saddle  & 2.149 &  2.0778 & 0 & 27.0123 \\
Lower LH saddle & -1.923 & -2.003  & 0 & 14.767 \\
Lower RH saddle & 1.923 & -2.003 & 0& 14.767 \\
\hline
\end{tabular}
\end{center}
\label{table:2}
\end{table}

\section{Results}
\label{results}

In this section we will analyze the structure of the periodic orbit dividing surfaces that correspond to the unstable periodic orbits of the upper right index-1 saddle (we will have similar results for the periodic orbit dividing surfaces that correspond to the other index-1 saddles as a result of the symmetry of the potential) in the energy surface (see subsection \ref{sub1}). Then we will present our results about the behavior of the trajectories that have initial conditions on these periodic orbit dividing surfaces and show how the periodic orbit dividing surfaces can detect the dynamical matching in a 3D Caldera-type system (see subsection \ref{sub2}).

\subsection{The construction and the structure of the periodic orbit dividing surfaces}
\label{sub1}

In this subsection we describe the computation and the structure of the periodic orbit dividing surfaces  for  a non-integrable Hamiltonian system with three degrees-of-freedom. For this purpose we use  our algorithm that we developed in previous papers (see \cite{katsanikas2021ds,katsanikas2021dsa}). We will use the first  algorithm of the construction of periodic orbit dividing surfaces that we proposed recently in \cite{katsanikas2021ds}. We will apply this method for a fixed value of energy ($E=28$) above the higher index-1 saddle.  

According to this algorithm we first compute one periodic orbit of the Lyapunov family of periodic orbits associated with the right higher index-1 saddle for a fixed value of energy ($E=28$). Then we follow the steps described below.

\begin{figure}
 \centering
 A)\includegraphics[scale=0.5]{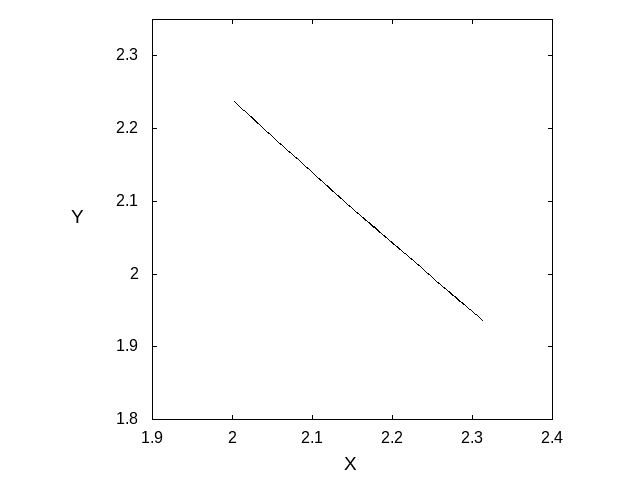}
 B)\includegraphics[scale=0.5]{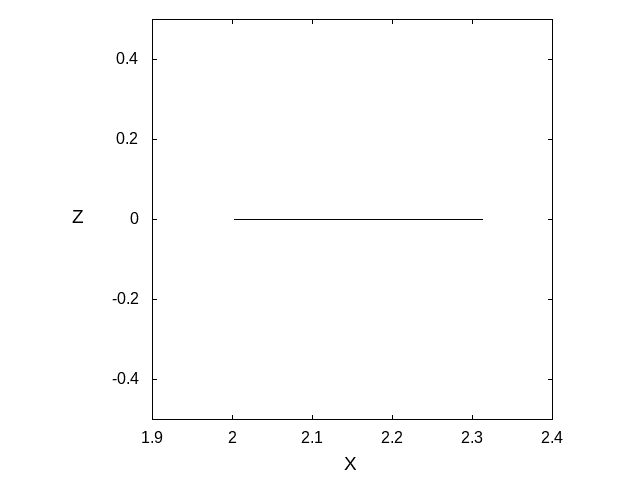}\\
 C)\includegraphics[scale=0.5]{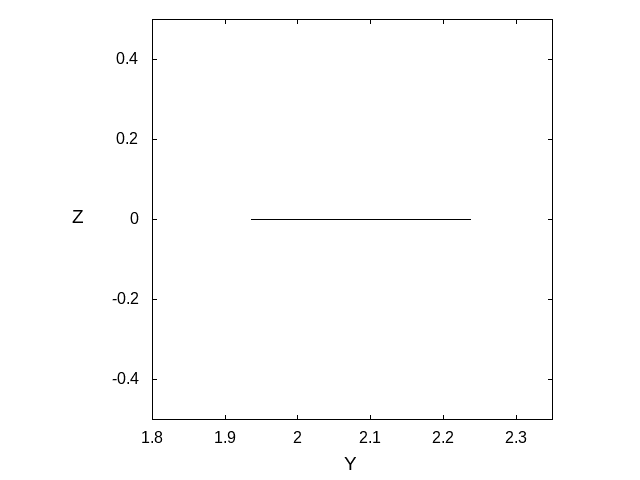}
 D) \includegraphics[scale=0.5]{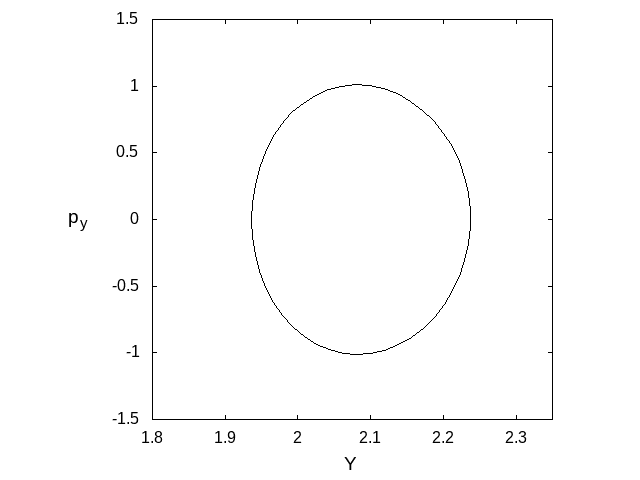}\\
\caption{The ($(x,y),(x,z),(y,z)$ and $(y,p_y$ projections of the periodic orbit of the Lyapunov family of the right higher index-1 saddle for $E=28$.  }
\label{3d-po}
\end{figure}

\begin{enumerate}
    \item We check if the periodic orbit is a closed curve in one of the 2D projections of the configuration space. This is crucial for choosing which version  of the algorithm of the construction of dividing surface is appropriate for our system (see \cite{katsanikas2021ds}).  In Fig. \ref{3d-po} we observe that the periodic orbit is not a closed curve in any 2D projection of the configuration space and this means that the second version  of the algorithm in \cite{katsanikas2021ds} is not valid. 
    
    \item Therefore we need to check if the first version of the algorithm is appropriate in our case. For this reason, we inspect if the  periodic orbit is a closed curve in any of the 2D projections of the phase space.  In Fig. \ref{3d-po} we see that this is true and this means that the first version of the algorithm is valid. The periodic orbit is represented by a closed curve in the $(y,p_y)$ subspace of the phase space. 
    
    \item We construct a torus that is produced from the Cartesian product of a circle in the $(x,y)$ plane with the projection of the periodic orbit in the $(y,p_y)$ subspace of the phase space. This Cartesian product gives us a two-dimensional torus in the $(x,y,p_y)$ space.  The radius of the circle, denoted $r$, is fixed at $r=0.25$.

    \item We construct a torus that is produced from the Cartesian product  of a circle in the $(y,z)$ plane with the torus that was constructed in the previous step. The result of this product is a three-dimensional torus in the $(x,y,z,p_y)$ space. The radius of the circle is fixed at $r=0.25$.
    
    \item We  find  for every point of the torus that was constructed in the previous step  the maximum and minimum value of the momentum $p_x$. After this we take values for $p_{x_i}$ $i=1,...n$ in the interval $[p_xmin,p_xmax]$. In this way, we add  a segment at the torus of the previous step increasing its dimensionality from three to four. In this step we used $n=5$ for our computations.  
    
    \item For every point of the four-dimensional torus that was constructed in the previous step we find the last momentum $p_z$ from the Hamiltonian.
    
    \end{enumerate}
    
\begin{figure}
 \centering
 \includegraphics[scale=0.73]{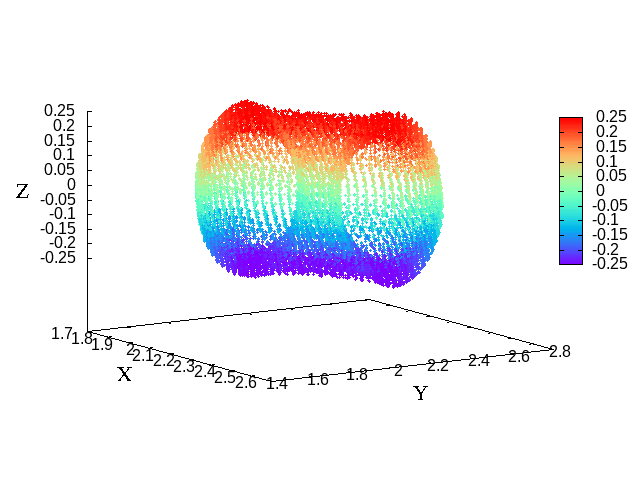}\\
\caption{The periodic orbit dividing surface for $E=28$ in the $(x,y,z)$ subspace of the phase space. The color indicates the third dimension. }
\label{3d-div1}
\end{figure}

\begin{figure}
 \centering
 A)\includegraphics[scale=0.73]{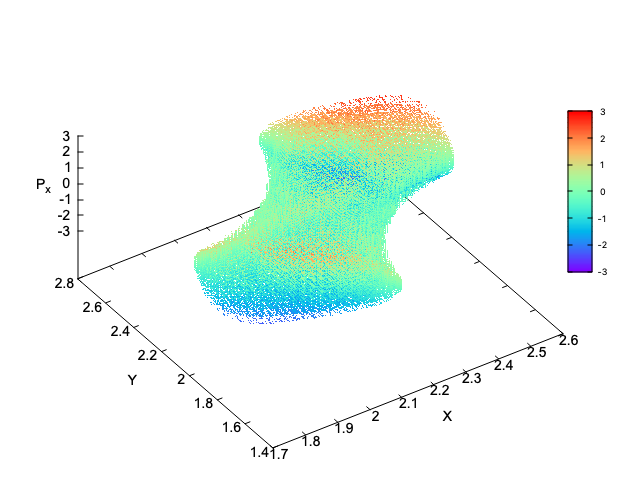}\\
 B) \includegraphics[scale=0.73]{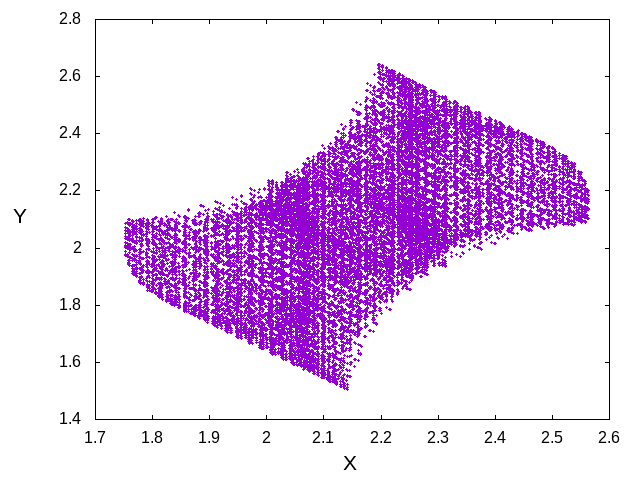}\\ 
\caption{Panel A:The periodic orbit dividing surface for $E=28$ in the $(x,y,p_x)$ subspace of the phase space. The color indicates the third dimension. Panel B: The 2D projection $(x,y)$ of the periodic orbit dividing surface for $E=28$.}
\label{3d-div2}
\end{figure}

\begin{figure}
 \centering
 \includegraphics[scale=0.73]{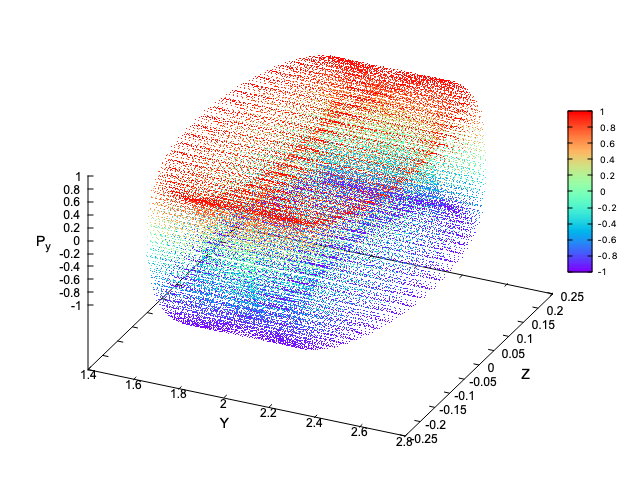}\\
\caption{The periodic orbit dividing surface for $E=28$ in the $(y,z,p_y)$ subspace of the phase space.The color indicates the third dimension.}
\label{3d-div3}
\end{figure}

   The structure of the periodic orbit dividing surfaces that were constructed from the  algorithm is depicted in Figs. \ref{3d-div1}, \ref{3d-div2} and \ref{3d-div3}. In order to see better the structure of these objects we increase the number of values of $p_x$ (from $n=5$ to  $n=21$) in the fifth step of the  algorithm. The periodic orbit dividing surface is a 4-dimensional torus in the 5-dimensional energy surface $(x,y,z,p_x,p_y)$ (see \cite{katsanikas2021ds}). This structure  is represented  by toroidal structures in the 3D projections $(x,y,z), (x,y,p_y),(x,z,p_y)$ and $(x,p_x,p_y)$ of the phase space that are similar to the structure of Fig. \ref{3d-div1}. Furthermore, the periodic orbit dividing surface has the morphology of a hyperboloid in the 3D projections $(x,y,p_x)$ and $(y,p_x,p_y)$ of the phase space (see for example the 3D projection $(x,y,p_x)$ in  panel A of Fig. \ref{3d-div2}). The hyperbolic behavior of this  structure can be seen better in some of the  2D projections of this object (see for example the projection $(x,y)$ in the panel B of Fig. \ref{3d-div2}). In the other 3D projections, the periodic orbit dividing surface has a boxy structure (see the boxy structure in Fig. \ref{3d-div3}).

\subsection{Dynamical Matching}
\label{sub2}

\begin{figure}
 \centering
 A)\includegraphics[scale=0.73]{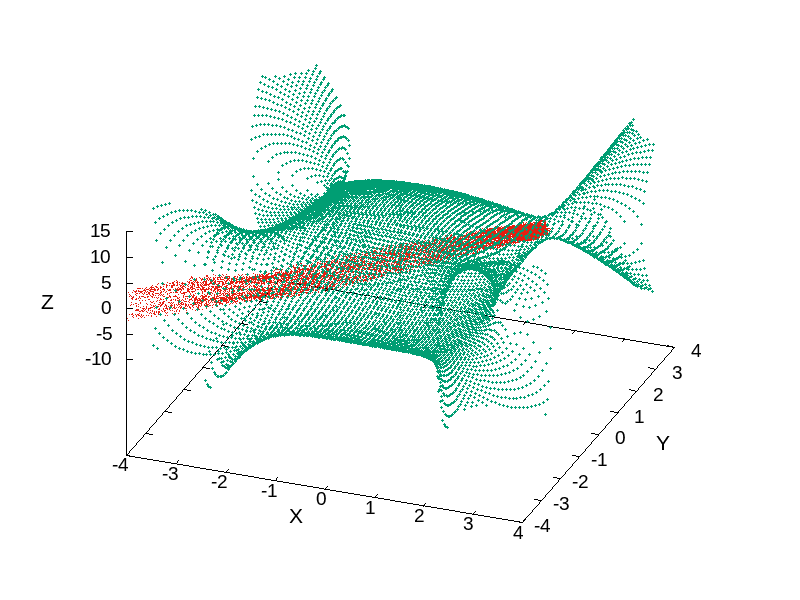}\\
 B) \includegraphics[scale=0.73]{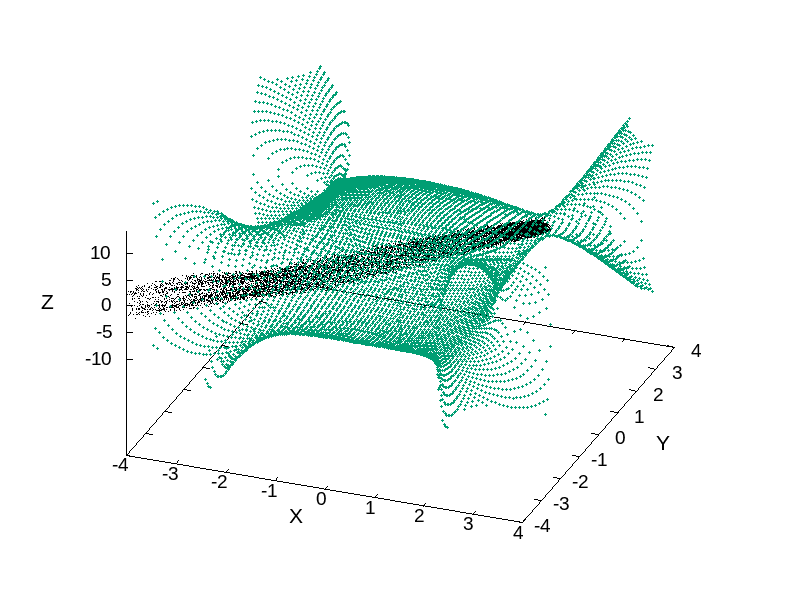}\\ 
\caption{Panel A: The evolution of the  trajectories (with red forward in time), that have initial conditions on the dividing surface associated with the periodic orbits of the upper right index-1 saddle and they move in the  direction of the reaction, in the configuration space $(x,y,z)$.   Panel B: The evolution of the  trajectories (with black backward in time), that have initial conditions on the dividing surface associated with the periodic orbits of the upper right index-1 saddle and they move in the  direction of the reaction, in the configuration space $(x,y,z)$. In all panels, we depict the zero velocity surface using a green color.}
\label{f1b1}
\end{figure}

\begin{figure}
 \centering
 A)\includegraphics[scale=0.73]{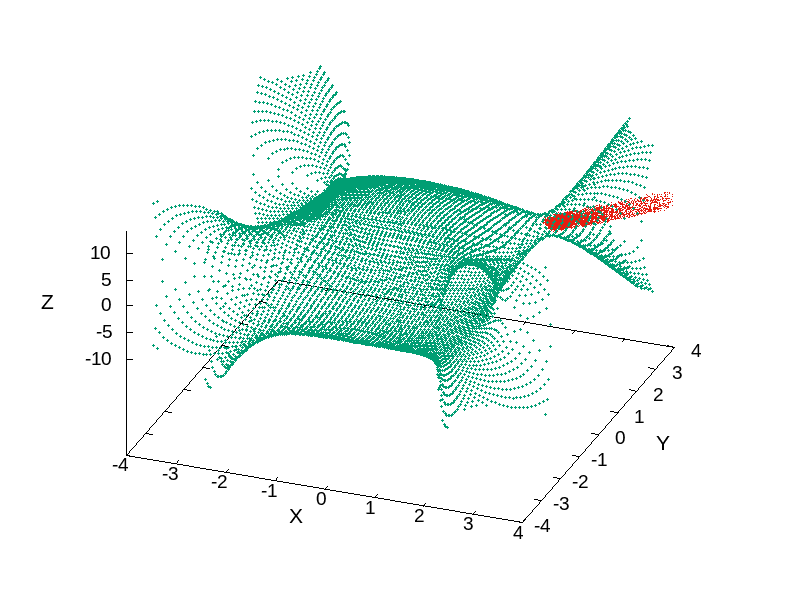}\\
 B) \includegraphics[scale=0.73]{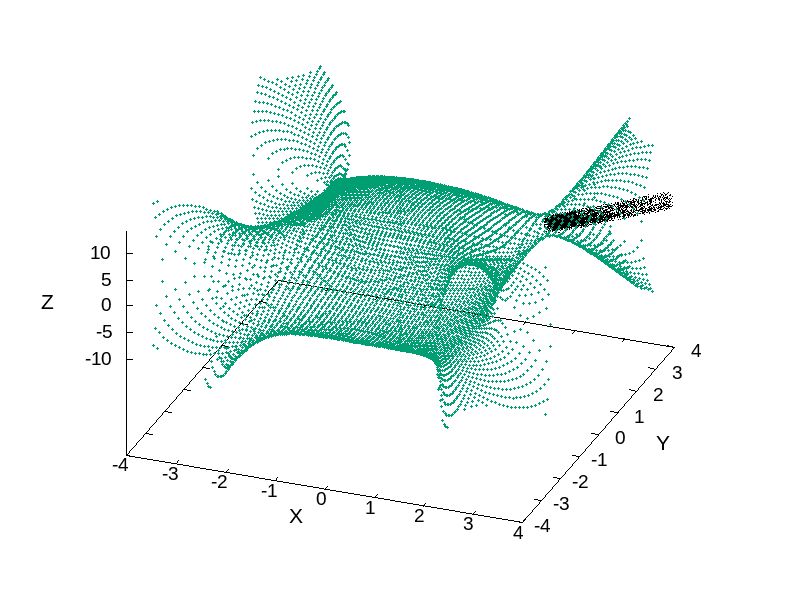}\\ 
\caption{Panel A: The evolution of the  trajectories (with red forward in time), that have initial conditions on the dividing surface associated with the periodic orbits of the upper right index-1 saddle and they move in the opposite direction of the reaction, in the configuration space $(x,y,z)$.   Panel B: The evolution of the  trajectories (with black backward in time), that have initial conditions on the dividing surface associated with the periodic orbits of the upper right index-1 saddle and they move in the opposite direction of the reaction, in the configuration space $(x,y,z)$. In all panels, we depict the zero velocity surface using green color.}
\label{f2b2}
\end{figure}

\begin{figure}
 \centering
 \includegraphics[scale=0.73]{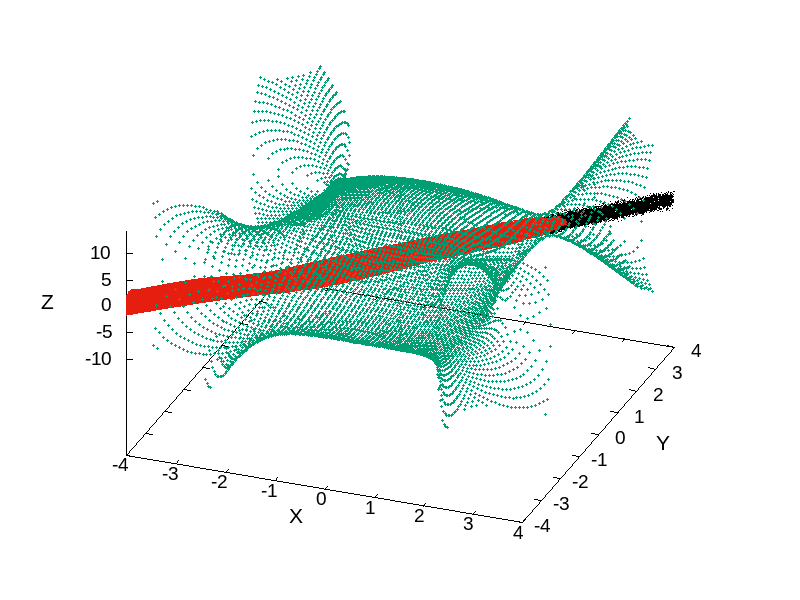}\\
\caption{The evolution of the  trajectories (with red forward in time), that have initial conditions on the dividing surface associated with the periodic orbits of the upper right index-1 saddle and they move in the  direction of the reaction, in the configuration space $(x,y,z)$.   The evolution of the  trajectories (with black backward in time), that have initial conditions on the dividing surface associated with the periodic orbits of the upper right index-1 saddle and they move in the opposite direction of the reaction, in the configuration space $(x,y,z)$. In all panels, we depict the zero velocity surface using a green color.}
\label{f1b2}
\end{figure}

\begin{figure}
 \centering
 \includegraphics[scale=0.73]{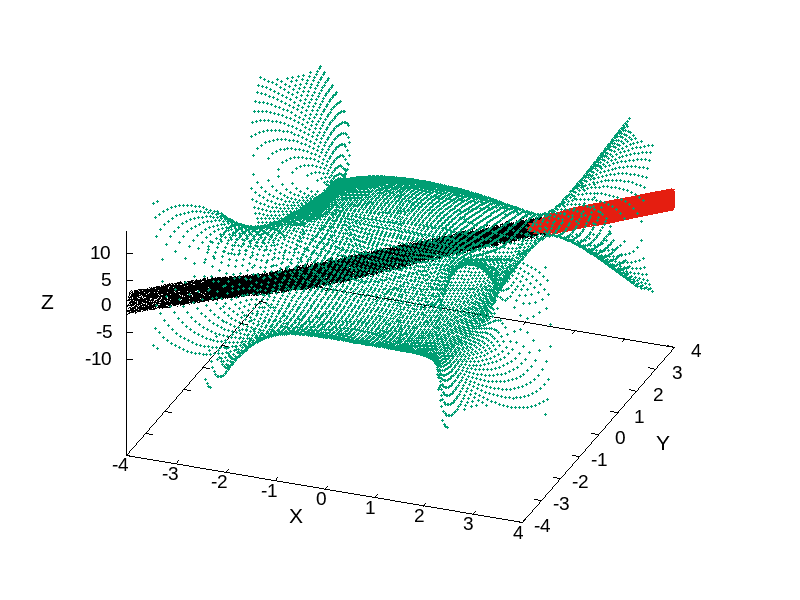}\\
\caption{The evolution of the  trajectories (with red forward in time), that have initial conditions on the dividing surface associated with the periodic orbits of the upper right index-1 saddle and they move in the opposite direction of the reaction, in the configuration space $(x,y,z)$.   The evolution of the  trajectories (with black backward in time), that have initial conditions on the dividing surface associated with the periodic orbits of the upper right index-1 saddle and they move in the  direction of the reaction, in the configuration space $(x,y,z)$. In all panels, we depict the zero velocity surface using a green color.}
\label{f2b1}
\end{figure}

In this subsection we study the behavior of the trajectories that have initial conditions on the periodic orbit dividing surface associated with the upper right index-1 saddle for $E=28$. We construct the periodic orbit dividing surface using the algorithm described in the previous subsection (for $r=0.25,n=5$ that corresponds to 21.178 initial conditions). Then we integrated the initial conditions on this periodic orbit dividing surface forward and backward in time for 7 time units (this time interval is enough for all trajectories to exit from the caldera). We classified the resulting trajectories into reactive and non-reactive trajectories and labelled the initial conditions on the dividing surface accordingly. The reactive trajectories have a direction towards the center of the caldera forward in time. On the contrary, the non-reactive trajectories become unbounded forward in time. We encountered four types of trajectory behavior:

\begin{enumerate}
    \item {\bf First type:} For this type, the trajectories (7176 trajectories $\approx 33.88\%$ of the initial trajectories on the dividing surface), that start from the periodic orbit dividing surface associated with the right upper index-1 saddle, go straight across the caldera (see Fig. \ref{f1b1}) and exit through the region of the opposite index-1 saddle. This happens if we integrate the initial conditions on the dividing surface forward and backward. This means that the trajectories go initially to the direction of the reaction (that is the region of the lower index-1 saddles) and exit through only from the region of the opposite lower index-1 saddle. These trajectories belong to the reactive part of this periodic orbit dividing surface and correspond to   $\approx 33.88\%$ of the initial trajectories on the dividing surface.
    
    \item {\bf Second type:} For this type, if we integrate (forward or backward) the initial conditions on the periodic orbit dividing surface associated with the right upper index-1 saddle, the resulting trajectories go directly to   infinity (see Fig. \ref{f2b2}). These trajectories   (7312 trajectories, $\approx 34.53\%$ of the initial trajectories on the dividing surface) belong to the non-reactive part of this periodic orbit dividing surface.
    
    \item {\bf Third type:}  For this type, the trajectories (that start from the periodic orbit dividing surface associated with the right upper index-1 saddle) go straight across the caldera and they exit through the opposite lower saddle under forward integration (see Fig. \ref{f1b2}). These trajectories go directly to  infinity backward in time (see Fig. \ref{f1b2}). This means that these trajectories (3314 trajectories $\approx 15.65\%$ of the initial trajectories on the dividing surface) belong to the reactive part of this periodic orbit dividing surface. 
    
    \item {\bf fourth type:} For this type, the trajectories (with initial conditions on the periodic orbit dividing surface associated with the right upper index-1 saddle) go directly to the infinity forward in time and go straight across the caldera, and exit through the lower opposite saddle backward in time (see Fig. \ref{f2b1}). In this type, the trajectories (3376 trajectories $\approx 15.94\%$ of the initial trajectories on the dividing surface) do not belong to the reactive part of the dividing surface because they move in the opposite direction forward in time. 
\end{enumerate}

We observe that the first and third types of trajectories correspond to the reactive part ($\approx 49.53\%$ of the initial conditions on the dividing surface) of the dividing surface and the second and fourth types of trajectories correspond to the non-reactive part of the dividing surface. All types of the trajectory behavior of the reactive part correspond to the phenomenon of dynamical matching since for these types all trajectories exit through the opposite lower saddle forward in time. 

\begin{figure}
 \centering
 A)\includegraphics[scale=0.73]{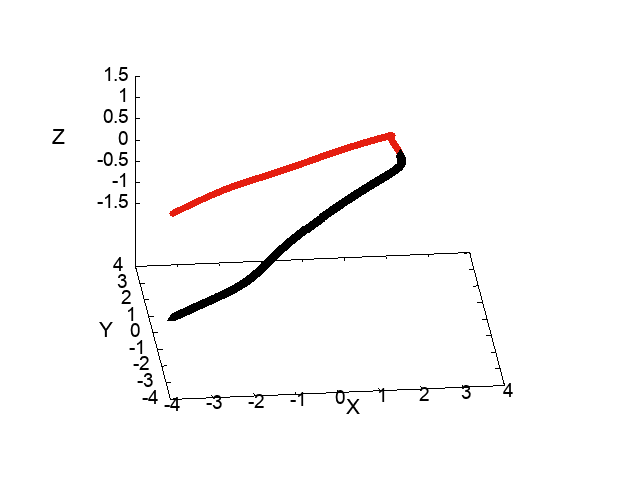}\\
 B)\includegraphics[scale=0.73]{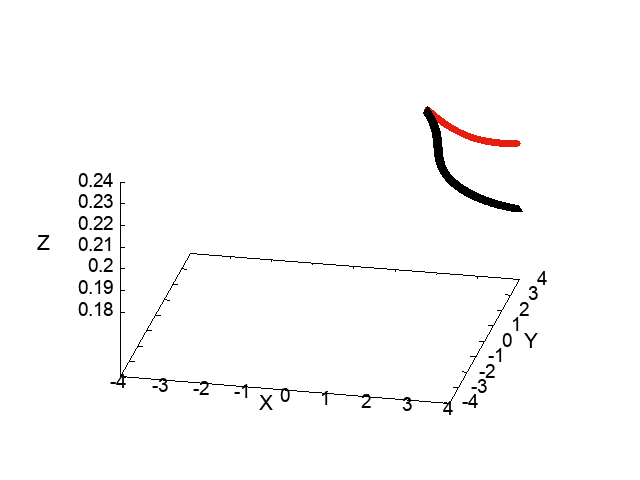}\\
\caption{A) One first type trajectory (with red and black color in forward and backward in time respectively). B) One second type trajectory (with red and black color in forward and backward in time respectively).}
\label{fb-example}
\end{figure}

Furthermore, we note that the first and second type of trajectory behavior includes trajectories that you can find only in the 3D caldera and not in the 2D caldera. This happens because in these cases the trajectories move in the same direction (in the direction of reaction or in the opposite direction) forward and backward in time, following different paths in the third dimension. These trajectories (Fig.\ref{fb-example})  form a horseshoe embedded in the 3D configuration space. The one branch of this horseshoe corresponds to the forward evolution in time and the other to the backward evolution in time. This explains why the trajectories move in the same direction forward and backward in time. 

The trajectory behavior of Fig. \ref{fb-example} resembles the trajectory behavior that we encounter in lobes between homoclinic intersections. For example, if we had a trajectory in a lobe between homoclinic intersections of the stable and unstable manifold of the NHIM that is associated with the lower-left index-1 saddle, we would have the same trajectory behavior with that of the panel A of Fig. \ref{fb-example}. Similarly, if we had a trajectory in a lobe between homoclinic intersections of the stable and unstable manifolds of the NHIM associated with the upper right index-1 saddle, we would have the same trajectory behavior as that of the panel B of Fig. \ref{fb-example}. The Normally Hyperbolic Invariant Manifold -NHIM (see for example \cite{wiggins2016}) is a 3D geometrical object in the phase space of a Hamiltonian system with three degrees of freedom. The invariant manifolds of the NHIMs are 4-dimensional objects in the phase space of a Hamiltonian system with three degrees of freedom. The 2D slices of the invariant manifolds of NHIMs will be 1-dimensional objects and will be presented as lines or curves in these slices.

\begin{figure}
 \centering
 A)\includegraphics[scale=0.56]{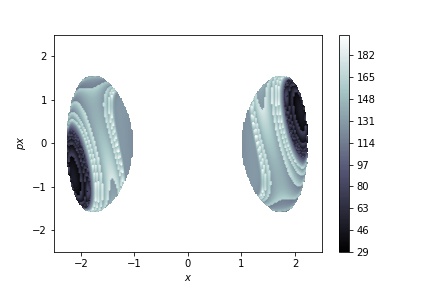}
 B)\includegraphics[scale=0.56]{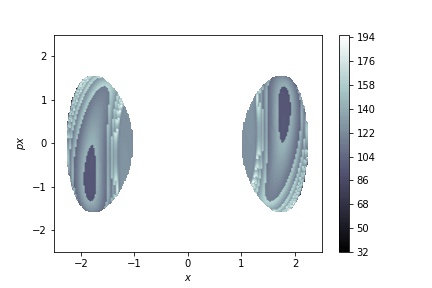}\\
 C)\includegraphics[scale=0.56]{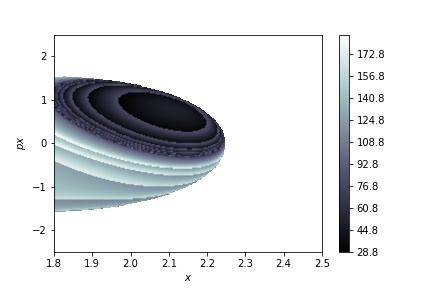}
 D)\includegraphics[scale=0.56]{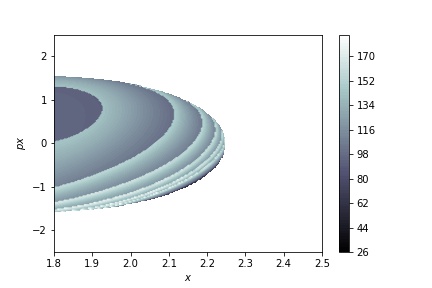}\\
\caption{Computation of variable-time LDs in the 2D slide $y= 1.946617848594, z= 0.237764129074, p_z= 0.912234902767$ using $\tau = 2$ and p = 1/2. A) Forward integration LDs;
B) Backward integration LDs; C) Forward integration for the interval $1.8<x<2.5$. D) Backward integration for the interval  $1.8<x<2.5$. }
\label{lds}
\end{figure}

\begin{figure}
 \centering
 A)\includegraphics[scale=0.65]{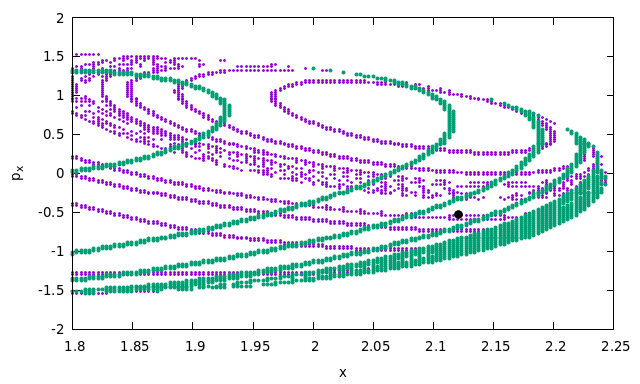}\\
\caption{The stable (violet) and unstable (green) invariant manifolds extracted (in the 2D slice $y= 1.946617848594, z= 0.237764129074, p_z= 0.912234902767$) from the gradient of the Lagrangian descriptors for the interval $1.8<x<2.5$. We depict using black color the initial condition of a first type trajectory that is shown in Fig. \ref{fb-example}.    }
\label{lds-1}
\end{figure}

In order to check the hypothesis of the previous paragraph for the origin of the trajectory behavior that is shown in Fig. \ref{fb-example}, we will use the method of  Lagrangian descriptors (LDs) (see the books \cite{Agaoglou2019,ldbook2020} and the references therein). The method of Lagrangian descriptors (LDs) is a trajectory-based scalar diagnostic tool that can be used to probe 2D slices of the phase space. We have used this method for analyzing the  2D Caldera potential energy surface recently (see \cite{katsanikas2020a,katsanikas2020b} -  the reader can find details in these references about the computation of LDs in a caldera-type potential). We will apply the same method in 2D slices of the phase space as we did for the 2D caldera potential energy surface (with the only difference that now we will integrate six equations of motion instead of four). In this paper we checked if the initial conditions of the trajectories of the first and second types trajectories belong in lobes between the stable and unstable invariant manifolds of NHIMs (that are computed using LDs). We present as an example the initial condition of the first type trajectory that is depicted in panel A of Fig. \ref{fb-example}. This trajectory has as initial condition  $x=  2.120469667800,y= 1.946617848594, z= 0.237764129074, p_x= -0.526679066677, p_y= 0.957219684607, p_z= 0.912234902767$. We will compute the LDs in the 2D slice $(x,p_x)$ space with  $y= 1.946617848594, z= 0.237764129074, p_z= 0.912234902767$ (computing $p_y$ from the Hamiltonian). We computed the LDs, using forward (for stable invariant manifolds) and backward (for unstable invariant manifolds) integration, in this slice (see the panels A and B of Fig.\ref{lds}). Then we restricted our analysis in a smaller area (see panels C and D of Fig. \ref{lds}) in which we have the initial condition of the trajectory in panel A of Fig. \ref{fb-example}. In order to visualize better the lobes between the intersections of the invariant manifolds of the NHIMs we extract the invariant manifolds from the gradient of LDs (see for details \cite{ldbook2020,katsanikas2020a}) and we plotted them together in Fig. \ref{lds-1}. We observe that the initial condition of the trajectory of panel A of Fig.\ref{fb-example} is in the lobe between the intersections of the stable and unstable invariant manifolds of a NHIM. As we can see in Fig. \ref{fb-example} this trajectory goes (forward and backward) to the region of the lower  left index-1 saddle. This means that the lobe, where the initial condition of this  trajectory is inside, is between the homoclinic intersections of the stable and unstable invariant manifolds of the NHIM associated with  the lower left index-1 saddle. Similarly, we can find that all the initial conditions of the first and second type trajectories are inside the lobes of homoclinic intersections of the invariant manifolds of the NHIM associated with the lower left index-1 saddle and of the NHIM associated with the the upper right index-1 saddle, respectively.

\section{conclusions}
\label{conc}

In this paper, we computed  the periodic orbit dividing surfaces in a 3D non-integrable Hamiltonian system. We studied the structure of these dividing surfaces and found that they have a Hyperboloid or a toroidal structure that is represented as a toroidal structure or a Hyperboloid or a box structure in 3D projections of the phase space. This result is in agreement with the structure of the same surfaces that are studied in the case of an integrable Hamiltonian system with three degrees of freedom (see \cite{katsanikas2021ds}). The only difference is that the periodic orbit dividing surfaces are not so smooth as in the integrable case of a Hamiltonian system with three degrees of freedom.

Using the periodic orbit dividing surfaces we  distinguished the reactive and non-reactive parts of a 3D Caldera potential energy surface. We found four types of trajectory behaviors associated with the upper index-1 saddles (trajectories that start from the periodic orbit dividing surfaces associated with the unstable periodic orbits of the upper index-1 saddles). The first and third types correspond to the reactive part and the second and fourth types to the non-reactive part. The reactive part is about $49.53\%$ of the trajectories that start from the periodic orbit dividing surface associated with the upper index-1 saddles. All trajectories of the reactive part go straight across the caldera and they exit through the region of the opposite lower index-1 saddle. This is the phenomenon of dynamical matching, that was detected for the first time in a 3D potential energy surface. The first and second types of  trajectories are a consequence of homoclinic intersections of the invariant manifolds  of the NHIM associated with the lower index-1 saddles and the opposite upper index-1 saddles, respectively. 

\section*{acknowledgements}
The authors acknowledge the financial support provided by the EPSRC Grant No. EP/P021123/1.

\clearpage
\bibliography{caldera2c}

\end{document}